\documentclass[useAMS,usenatbib]{mn2e}
\usepackage{graphicx}
\def\aj{AJ}%
\def\apj{ApJ}%
\def\apjl{ApJ}%
\def\aap{A\&A}%
\def\mnras{MNRAS}%
\title[The formation of brown dwarfs]
{Brown dwarf formation by gravitational fragmentation of massive, extended protostellar discs}
\author[D. Stamatellos, D.~A. Hubber \& A.~P. Whitworth]
{Dimitris Stamatellos\thanks{E-mail:D.Stamatellos@astro.cf.ac.uk}, David A. Hubber and Anthony~P. Whitworth \\ 
School of Physics \& Astronomy,Cardiff University, Cardiff, CF24 3AA, Wales, UK}

\begin{document}

\date{Accepted 2007 August 17. Received 2007 July 31; in original form 2007 July 5}

\pagerange{\pageref{firstpage}--\pageref{lastpage}} \pubyear{2007}

\maketitle

\label{firstpage}

\begin{abstract}
We suggest that low-mass hydrogen-burning stars like the Sun should sometimes form with massive extended discs; and we show, by means of radiation hydrodynamic simulations, that the outer parts of such discs ($R\ga 100\,{\rm AU}$) are likely to fragment on a dynamical timescale ($10^3$ to $10^4\,{\rm yr}$), forming low-mass companions: principally brown dwarfs (BDs), but also very low-mass hydrogen-burning stars and planetary-mass objects. A few of the BDs formed in this way remain attached to the primary star, orbiting at large radii. The majority are released into the field, by interactions amongst themselves; in so doing they acquire only a low velocity dispersion ($\la 2\,{\rm km}\,{\rm s}^{-1}$), and therefore they usually retain small discs, capable of registering an infrared excess and sustaining accretion. Some BDs form close BD/BD binaries, and these binaries can survive ejection into the field. This BD formation mechanism appears to avoid some of the problems associated with the `embryo ejection' scenario, and to answer some of the questions not yet answered by the `turbulent fragmentation' scenario.
\end{abstract}


\section{Introduction}

There are two conditions that must be fulfilled for a disc to fragment. (i) The disc must be gravitationally unstable, i.e. massive enough so that gravity can overcome thermal pressure and centrifugal support (Toomre 1964):
\begin{equation}
Q(R)\;=\;\frac{c(R)\,\kappa(R)}{\pi\,G\,\Sigma(R)}\;\la\;1\,,
\end{equation}
where $c$ is the sound speed, $\kappa$ the epicyclic frequency, and $\Sigma$ the surface density. (ii) A proto-fragment must radiate away the compressional energy delivered by condensation on a dynamical time-scale (Gammie 2001; Rice et al. 2003),
\begin{equation}
t_{_{\rm COOL}}\; <\;{\cal C}(\gamma)\,t_{_{\rm ORB}}\,,\hspace{1.0cm}0.5\la{\cal C}(\gamma)\la 2.0\,,
\end{equation}
where $t_{_{\rm ORB}}$ is the period and $\gamma$ is the adiabatic exponent. 

Whilst there is general agreement that the above criteria must be satisfied for a disc to fragment, there is disagreement as to whether real discs actually satisfy these criteria (e.g. Durisen et al. 2007). The debate has focused on the {\it inner} regions of discs around Sun-like stars, $R \sim 3\,{\rm to}\,30\,{\rm AU}$, with Boss (2004) and Mayer et al. (2007) claiming that convection provides sufficiently rapid cooling for discs to fragment, whereas Johnson \& Gammie (2003), Mej\'ia et al. (2005), Boley et al. (2006) and Nelson (2006), assert that the cooling is too slow to allow fragmentation. The latter point of view is corroborated by analytic studies which suggest that convection cannot provide the required cooling (Rafikov 2007), and that the inner parts of discs ($R<100\,{\rm AU}$) cannot cool fast enough to fragment (Rafikov 2005; Matzner \& Levin 2005; Whitworth \& Stamatellos 2006).

In the present paper we switch the focus to what happens in the {\it outer} regions of discs around Sun-like stars. Whitworth \& Stamatellos (2006) argue that at large radii ($R\ga 100\,{\rm AU}$) discs can cool fast enough to fragment, and that this leads to the formation of BDs. A few of these BDs then remain as distant companions to the primary star, but the majority are released into the field. We report here a simulation corroborating these predictions. In Section 2 we explain why massive, extended discs of the type we invoke might form quite frequently. In Section 3 we describe the numerical method used, in particular the treatment of the energy equation and associated radiative transfer effects. In Section 4 we describe the simulation. In Section 5 we discuss the results, and argue that this mechanism could be an important source of BDs.

\section{Massive extended discs}

In the simulation reported here, the central star has mass $M_\star=0.7\,{\rm M}_{\sun}$. Initially the disc has mass $M_{_{\rm D}}=0.5\,{\rm M}_{\sun}$, inner radius $R_{_{\rm IN}}=40\,{\rm AU}$, outer radius $R_{_{\rm OUT}}=400\,{\rm AU}$, surface density 
\begin{equation}
\Sigma(R)=\frac{0.01\,{\rm M}_{\sun}}{{\rm AU}^2}\,\left(\frac{R}{\rm AU}\right)^{-7/4}\,,
\end{equation}
temperature
\begin{equation}
T(R)=300\,{\rm K}\,\left(\frac{R}{\rm AU}\right)^{-1/2}+10~{\rm  K}\,,
\end{equation}
and hence approximately uniform initial $Q\sim 1.2$. Thus the disc is marginally stable. In order to facilitate comparison with other simulations (e.g. Boley et al. 2006; Boss 2004), we do not include irradiation of the disc by the central primary protostar. Therefore the gas is heated solely by viscous dissipation, compression and the background radiation field. The effect of stellar irradiation will be the subject of a subsequent paper based on on-going simulations.

Only a small number of massive extended discs have been observed (e.g. Eisner et al. 2005, 2006; Rodriguez et al. 2005), but we suggest that this is because their outer parts are rapidly dissipated by gravitational fragmentation -- rather than because they seldom form in the first place. For example, a $1.2\,{\rm M}_\odot$ prestellar core with ratio of rotational to gravitational energy $\beta\equiv{\cal R}/|\Omega|$ will -- if it collapses monolithically -- form a central primary protostar and a protostellar disc with outer radius $R \sim 400\,{\rm AU}\,(\beta/0.01)$; the observations of Goodman et al. (1993) indicate that many prestellar cores may have $\beta \sim 0.02$. Alternatively, if an existing $0.7\,{\rm M}_\odot$ protostar attempts to assimilate matter with specific angular momentum $h$, this matter is initially parked in an orbit at $R \sim 400\,{\rm AU}\,(h/5\times 10^{20}\,{\rm cm}^2\,{\rm s}^{-1})^2$; again, this is a rather modest specific angular momentum by protostellar standards. Our simulations suggest that the outer parts of a massive disc with $R\sim 400\,{\rm AU}$ fragment on a timescale $\sim 3000\,{\rm yr}$, so they are very short-lived. 

\section{Numerical method}

For the hydrodynamic part of the simulation ($0\,{\rm to}\,15\times 10^3\,{\rm yr}$), we use the SPH code {\sc dragon} (Goodwin et al. 2004), which invokes an octal tree (to compute gravity and find neighbours), adaptive smoothing lengths, multiple particle timesteps, time-dependent artificial viscosity, and a second-order Runge-Kutta integration scheme. The disc is represented by 150,000 SPH particles, each of which has exactly 50 neighbours; we have tested that these numbers are sufficient for convergence. The energy equation takes into account compressional heating, viscous heating, radiative heating from the background, and radiative cooling. The functions of state take into account the rotational and vibrational degrees of freedom of H$_2$, the dissociation of H$_2$, and the ionization of H$^{\rm o}$, He$^{\rm o}$ and He$^{\rm +}$. For the dust and gas opacity we use the parameterization  by Bell \& Lin (1994), which takes account of ice mantle melting, dust sublimation, molecular-line, H$^-$, bound-free, free-free and electron scattering contributions. The associated radiative transfer effects are treated by an approximate method based on the diffusion approximation (Stamatellos et al. 2007); this method has been extensively tested and performs well in both the optically thin and the optically thick regimes.

For the $N$-body part of the simulation, exploring the longterm evolution of the ensemble of BD condensations formed by fragmentation of the outer disc ($15\times 10^3\,{\rm to}\,300\times 10^3\,{\rm yr}$), we use a 4th-order Hermite integration scheme (Makino \& Aarseth 1992), with a conservative timestep criterion so that energy is conserved to better than $10^{-8}$.

\section{Simulation}

Fig. 1 shows column-density images of the disc at $t=1000,\,1500,\,2000,\,2500,\,...\,5000\,{\rm yr}$. From Fig. 1 we see that the disc quickly becomes gravitationally unstable, and breaks up from the inside out, first into spiral arms, and then into protostellar condensations. The first condensation forms at $3.3\times 10^3\,{\rm yr}$, and by $\sim 8\times 10^3\,{\rm yr}$ it has grown close to its final mass of $\sim 58\,{\rm M}_{_{\rm J}}$. This first condensation is followed until its central density reaches $10^{-2}\,{\rm g}\,{\rm cm}^{-3}$, when it is replaced with a sink. Subsequent condensations are replaced with sinks when their central density reaches $10^{-9}\,{\rm g}\,{\rm cm}^{-3}$; all sinks have radius $1\,{\rm AU}$. This ensures that proto-condensations which should bounce, or be sheared apart, or merge with one another, can do so; they are only converted into sinks when they are very strongly bound and have very small cross-sections.

Fig. 2 shows the evolution of $\bar{Q}(R)$ and $\bar{t}_{_{\rm COOL}}(R)/t_{_{\rm ORB}}(R)$ (where the averages are azimuthal). After $t=5\times 10^3\,{\rm yr}$, so much mass has been converted into condensations that the disc ceases to be Toomre unstable, anywhere, and no further condensations form. The cooling time is always sufficiently short to allow fragmentation in the outer part of the disc, $R\ga 100\,{\rm AU}$.

Fig. 3 shows how the masses of individual condensations increase with time; condensations are only identified when they are replaced with a sink. The condensations which form early tend to grow in mass, both by steady accretion, and -- occasionally -- by assimilating smaller proto-condensations. The condensations which form later tend to grow rather slowly thereafter. By the end of the hydrodynamic simulation at $t=15\times 10^3\,{\rm yr}$, there are 11 condensations and their masses are almost constant. Only 15\% of the initial disc mass remains, and most of this is in the inner region ($<100\,{\rm AU}$).

Fig. 4 shows the distance of each condensation from the central star, as a function of time, up to the end of the SPH simulation at $t=15\times 10^3\,{\rm yr}$. We see that 2 single condensations are ejected, and two close binaries are formed. The ejections are essentially gravitational slingshots. Most of the condensations which are not ejected are on quite eccentric orbits around the central star.

The subsequent dynamical evolution of these condensations is then followed for a further $285\times 10^3\,{\rm yr}$ (i.e. to $t=300\times 10^3\,{\rm yr}$) using the $N$-body code, and Fig. 5 shows the results. The same pattern continues, with condensations being ejected periodically -- including the two BD/BD binary systems -- and those that remain tending to be on highly eccentric orbits. At the end, only two condensations remain bound to the central star, an $86\,{\rm M}_{_{\rm J}}$ condensation orbiting at $8\,{\rm AU}$, and a $36\,{\rm M}_{_{\rm J}}$ condensation orbiting at $208\,{\rm AU}$. We note that the $86\,{\rm M}_{_{\rm J}}$ condensation has become tightly bound by interacting with the ejected $10\,{\rm M}_{_{\rm J}}$ condensation, hence the latter's exceptionally high ejection velocity.

Table 1 summarises the key properties of the condensations. The majority have brown dwarf or planetary masses. There are two BD/BD binaries, both with separations of $\sim3\,{\rm AU}$. If we count these two binaries as single systems, then the ejected systems have velocities of $1.1,\,1.7,\,1.9,\,2.1,\,2.4,\,3.5,\,{\rm and}\,12.7\,{\rm km}\,{\rm s}^{-1}$. The BDs that are ejected are associated with discs of mass on the order of $10\,{\rm M}_{_{\rm J}}$. 

\begin{figure*}
\centerline{
\includegraphics[width=5cm,angle=-90]{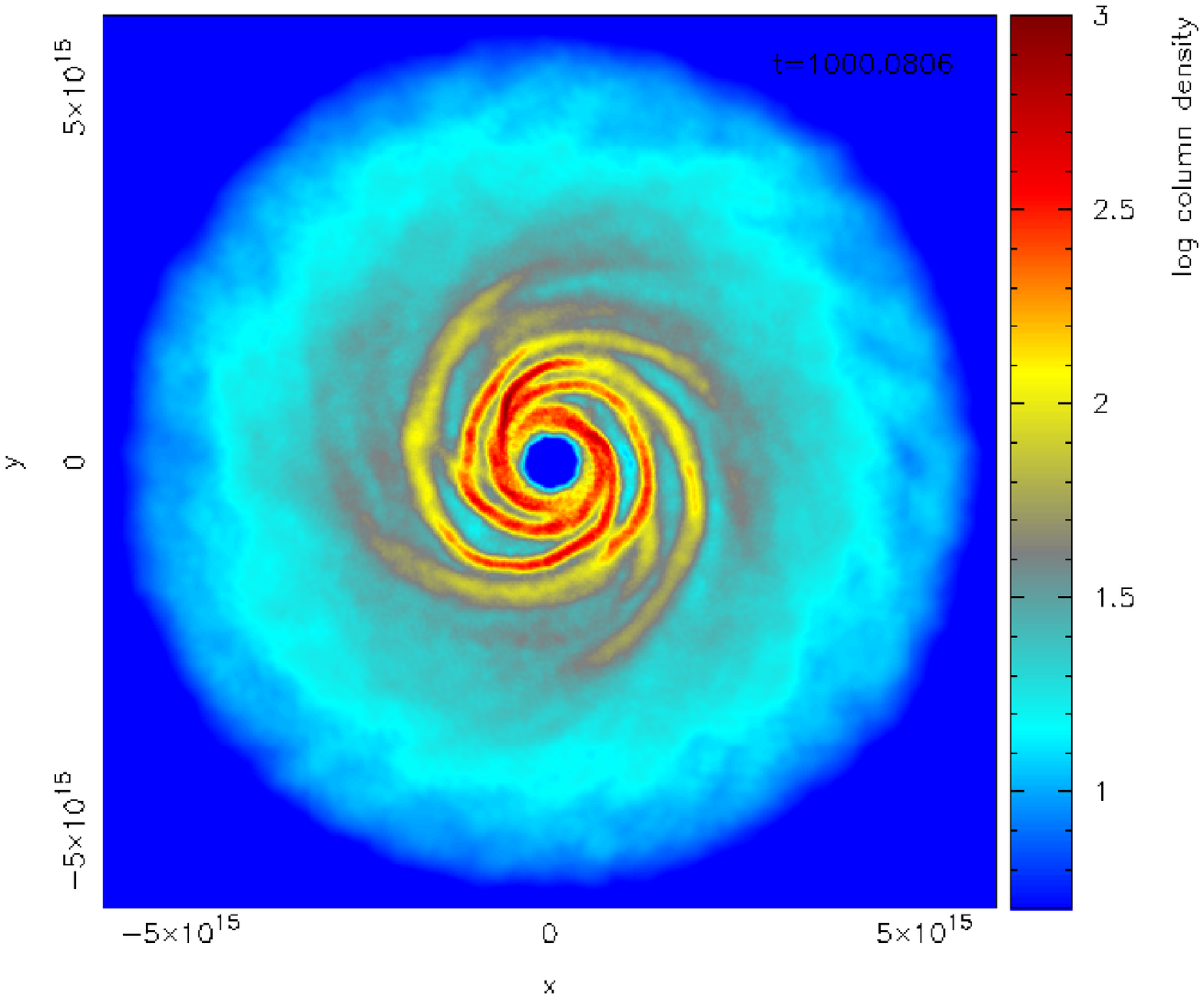}
\includegraphics[width=5cm,angle=-90]{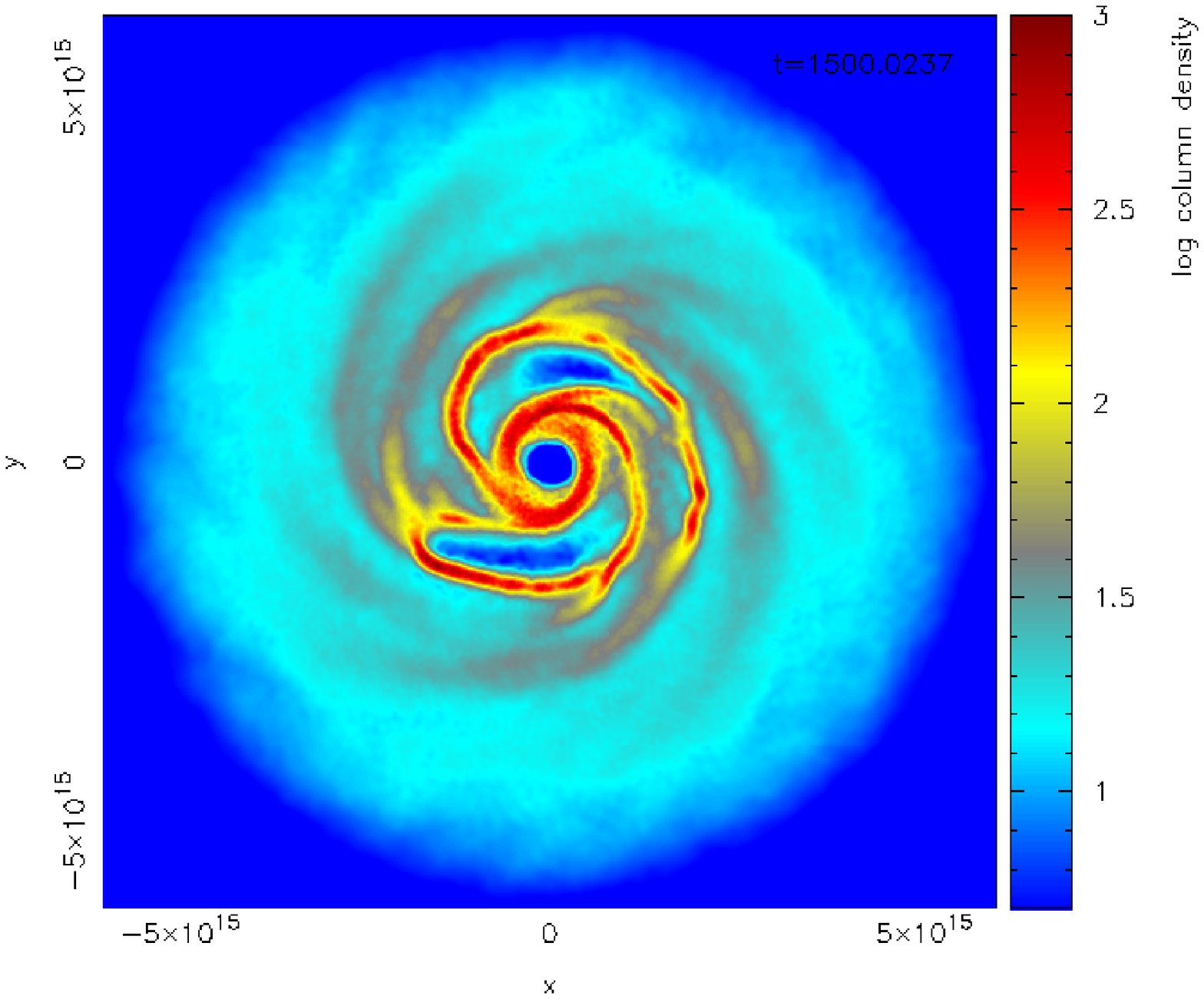}
\includegraphics[width=5cm,angle=-90]{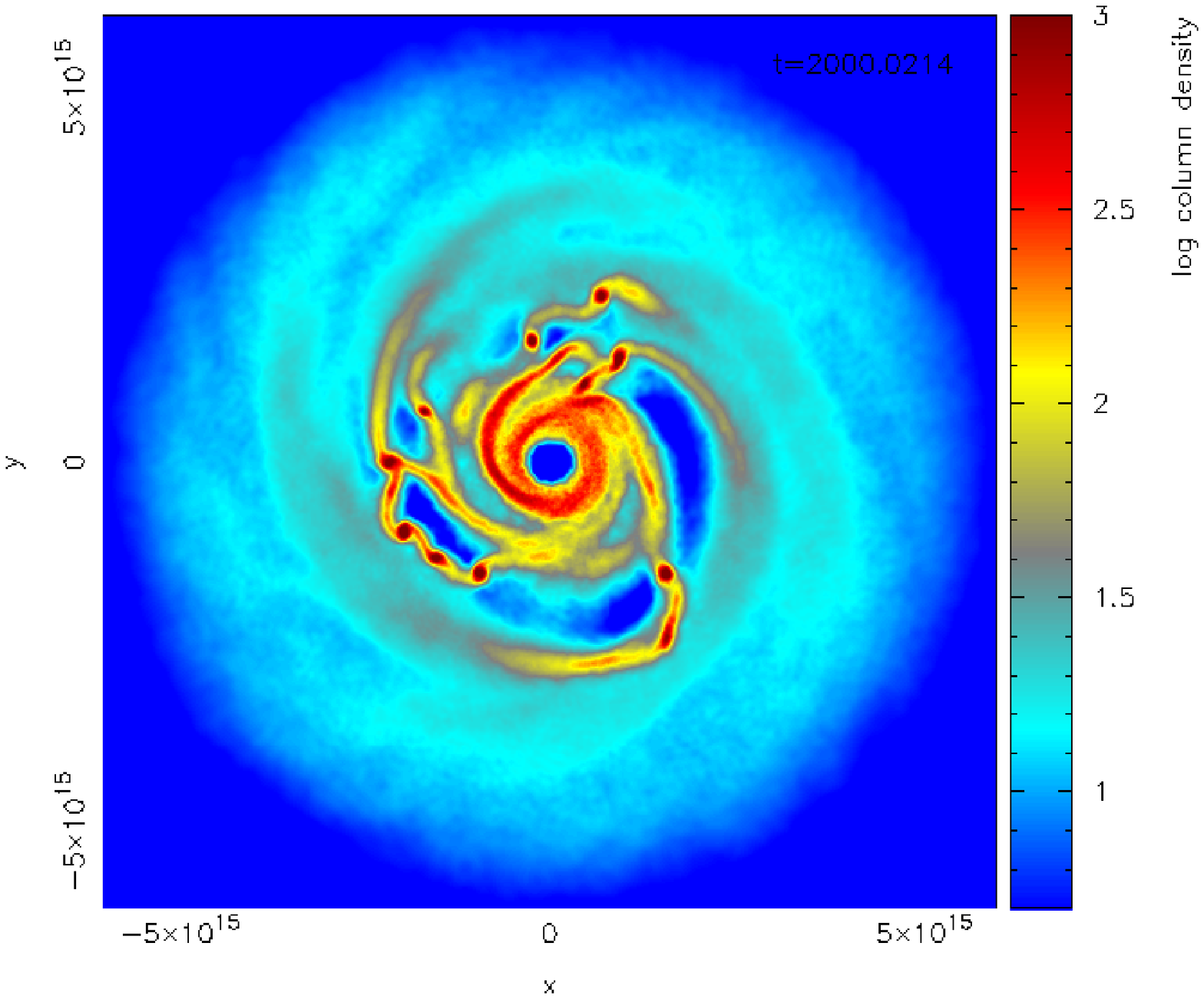}}
\centerline{
\includegraphics[width=5cm,angle=-90]{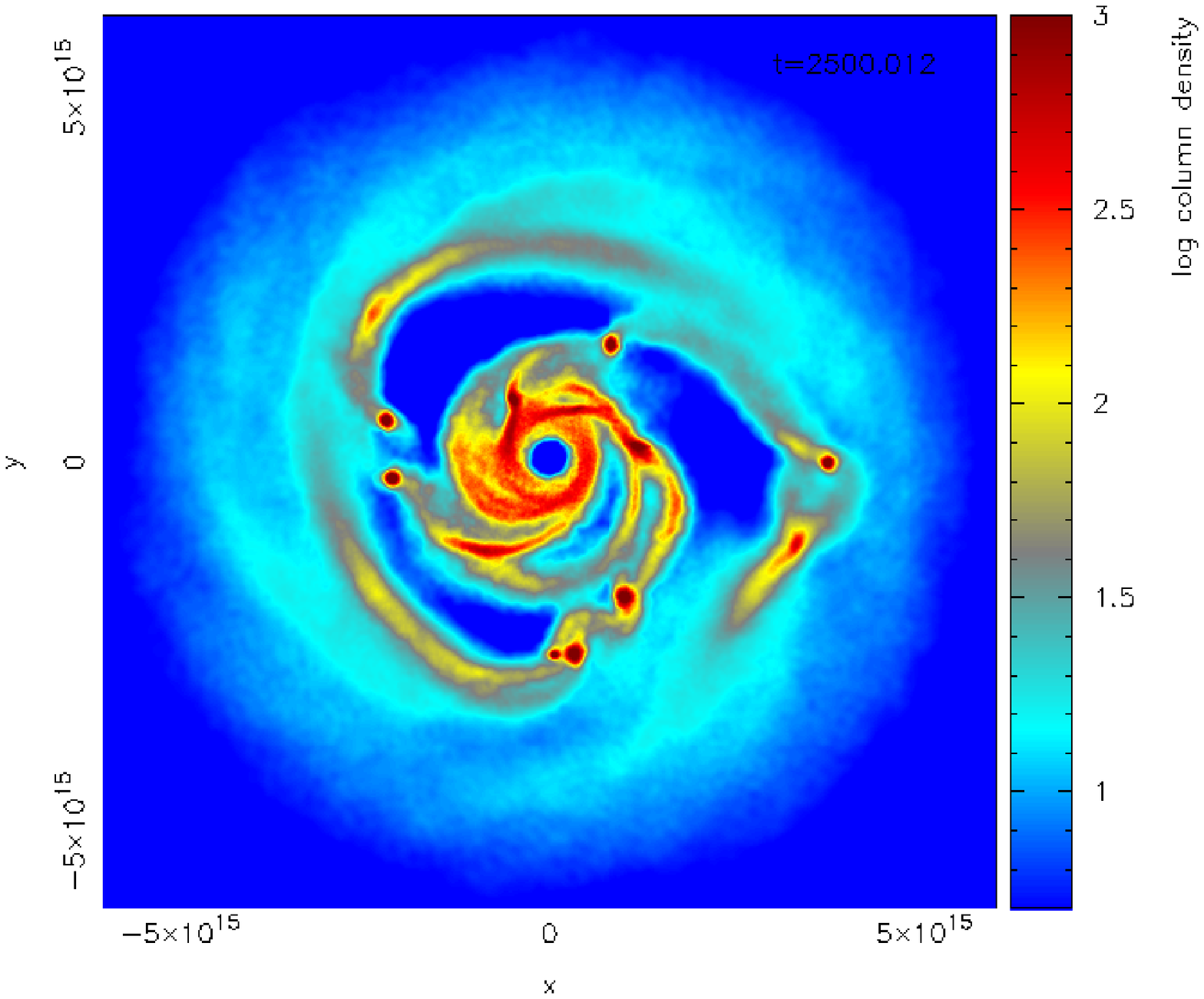}
\includegraphics[width=5cm,angle=-90]{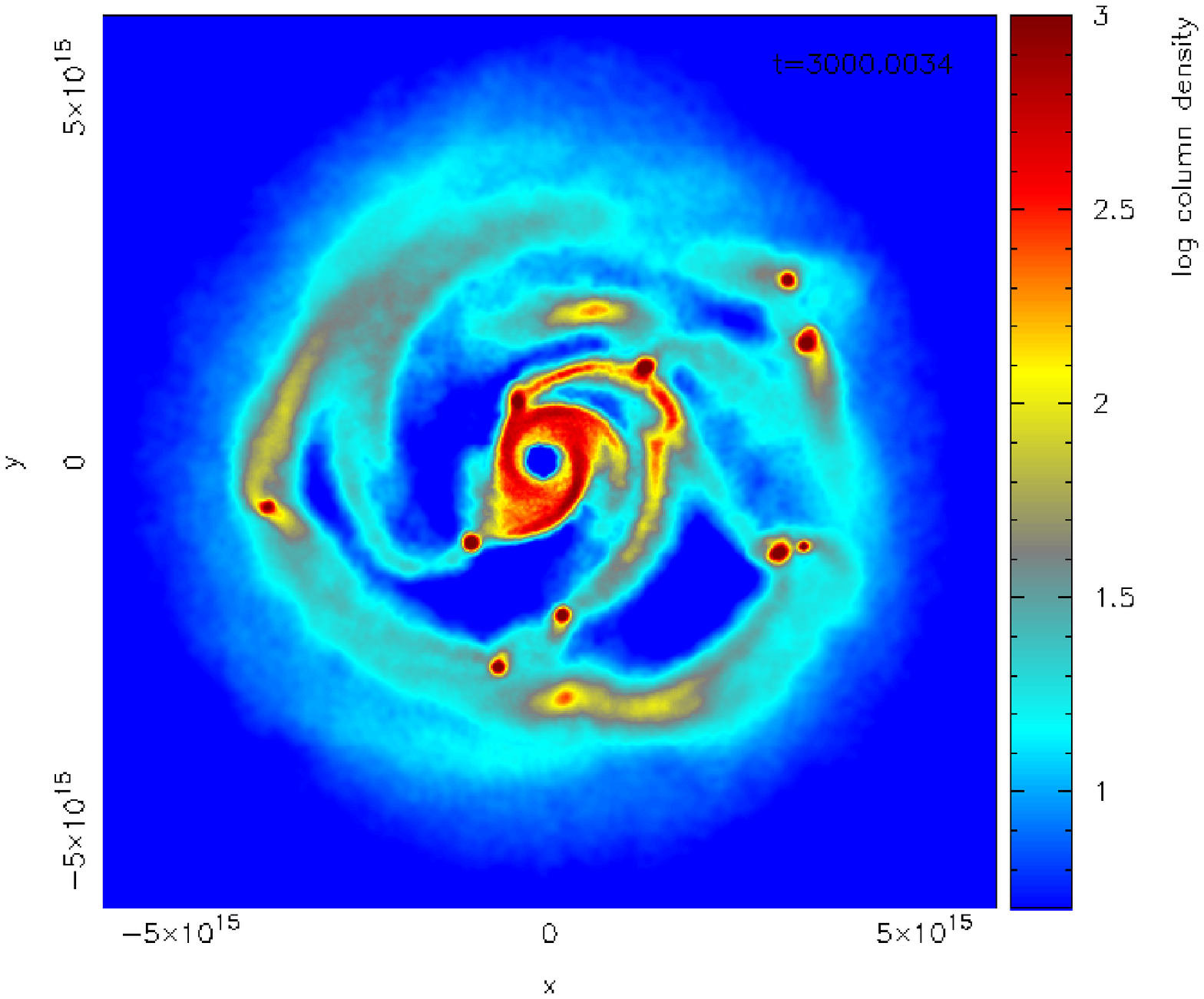}
\includegraphics[width=5cm,angle=-90]{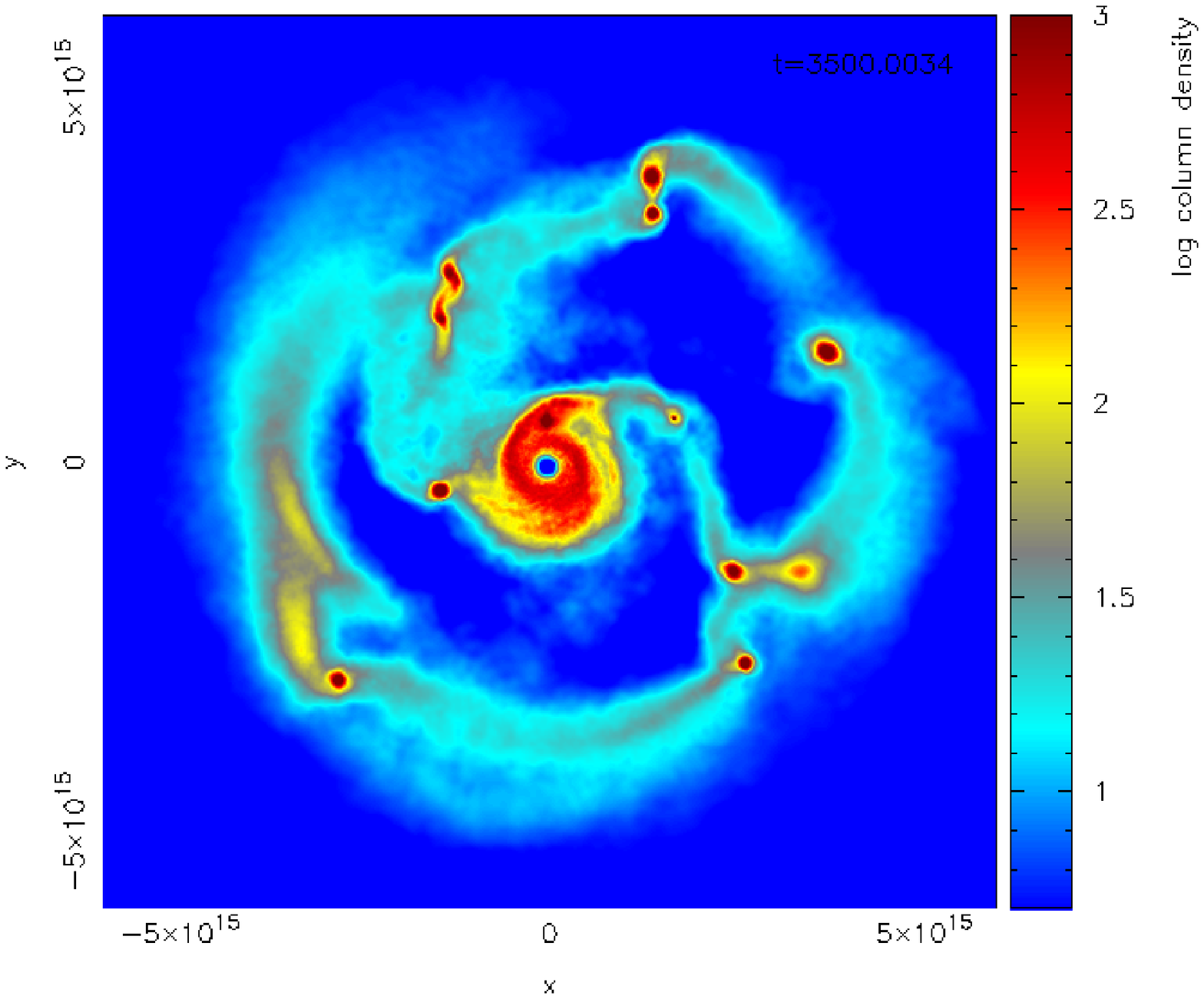}}
\centerline{
\includegraphics[width=5cm,angle=-90]{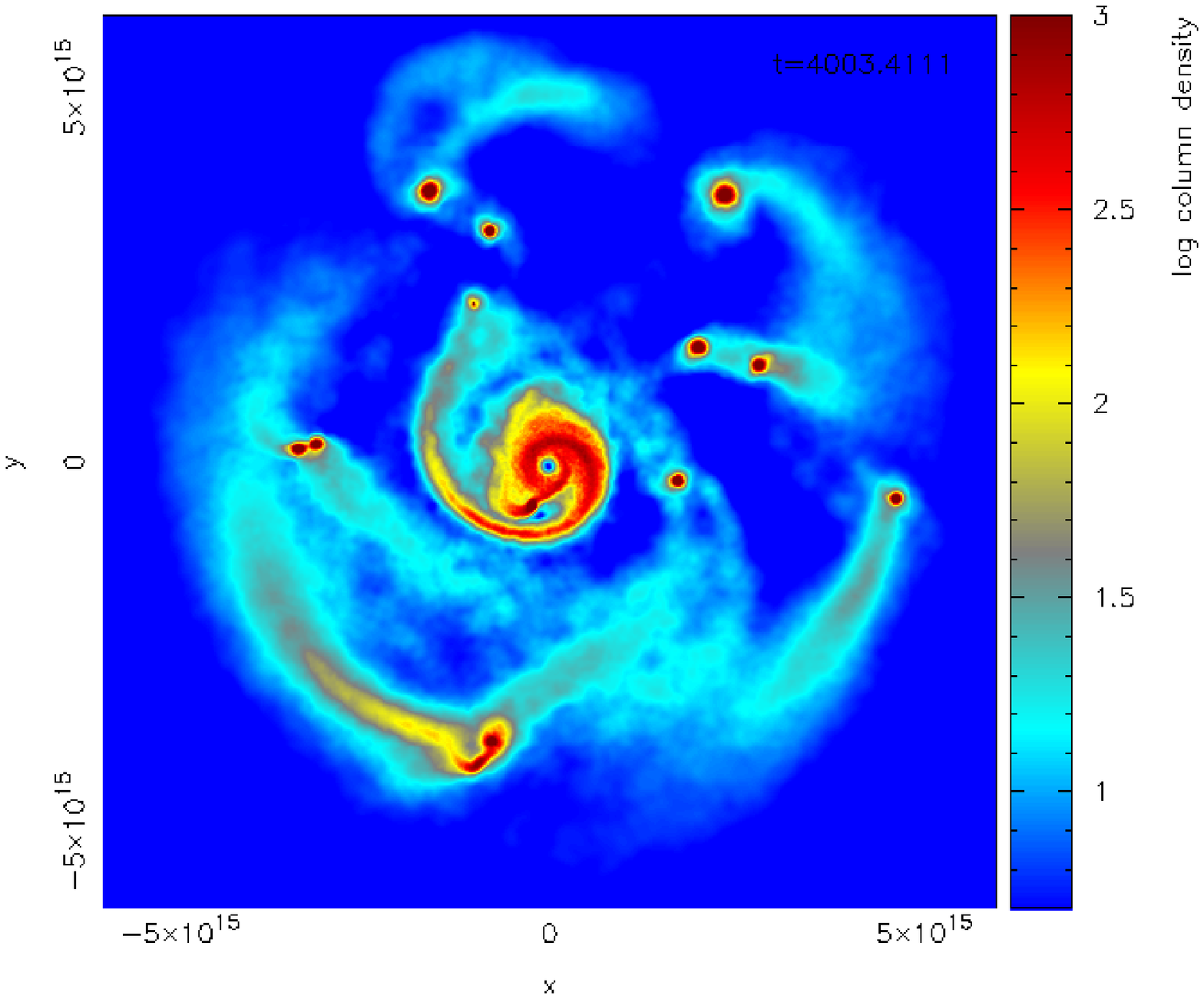}
\includegraphics[width=5cm,angle=-90]{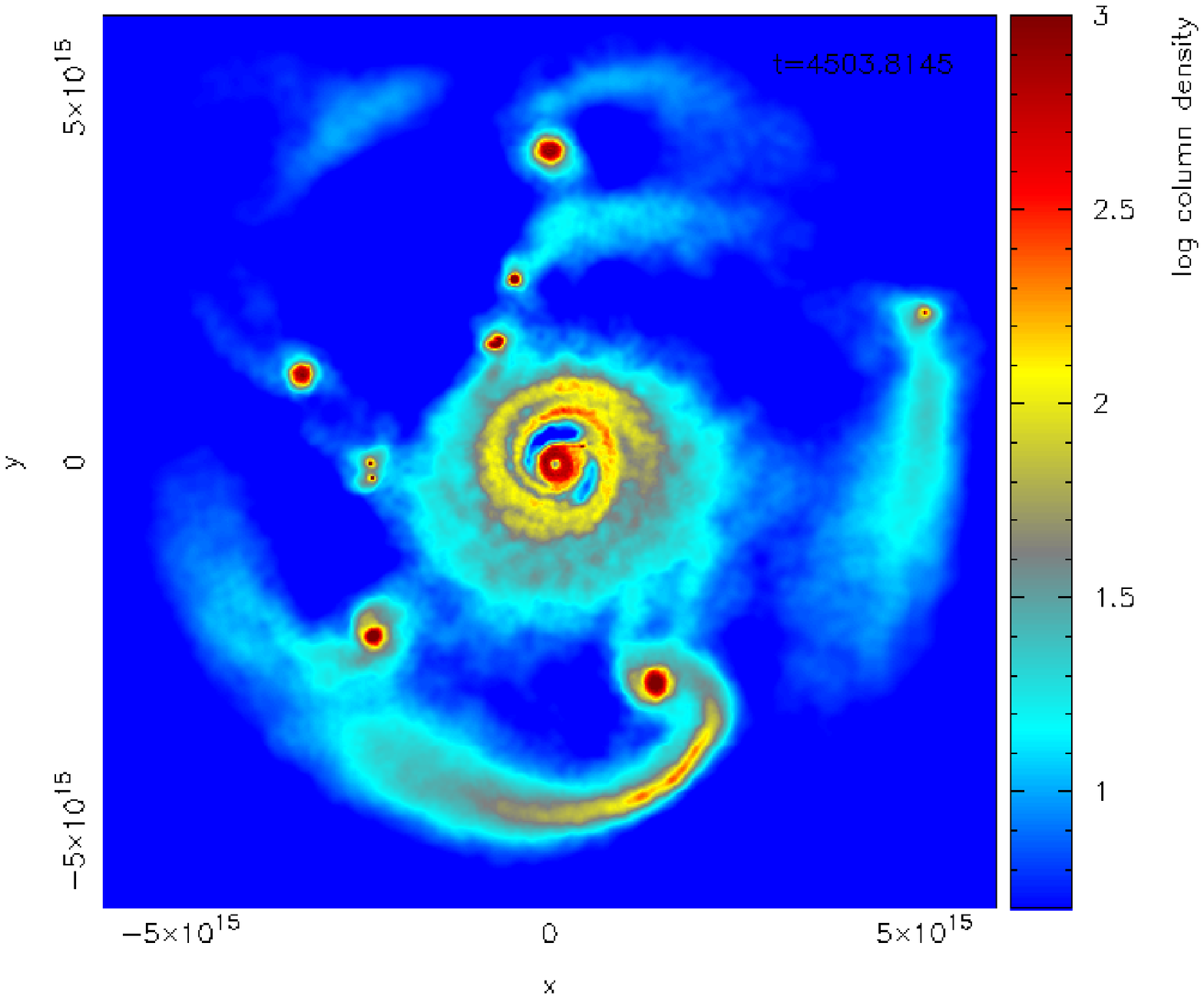}
\includegraphics[width=5cm,angle=-90]{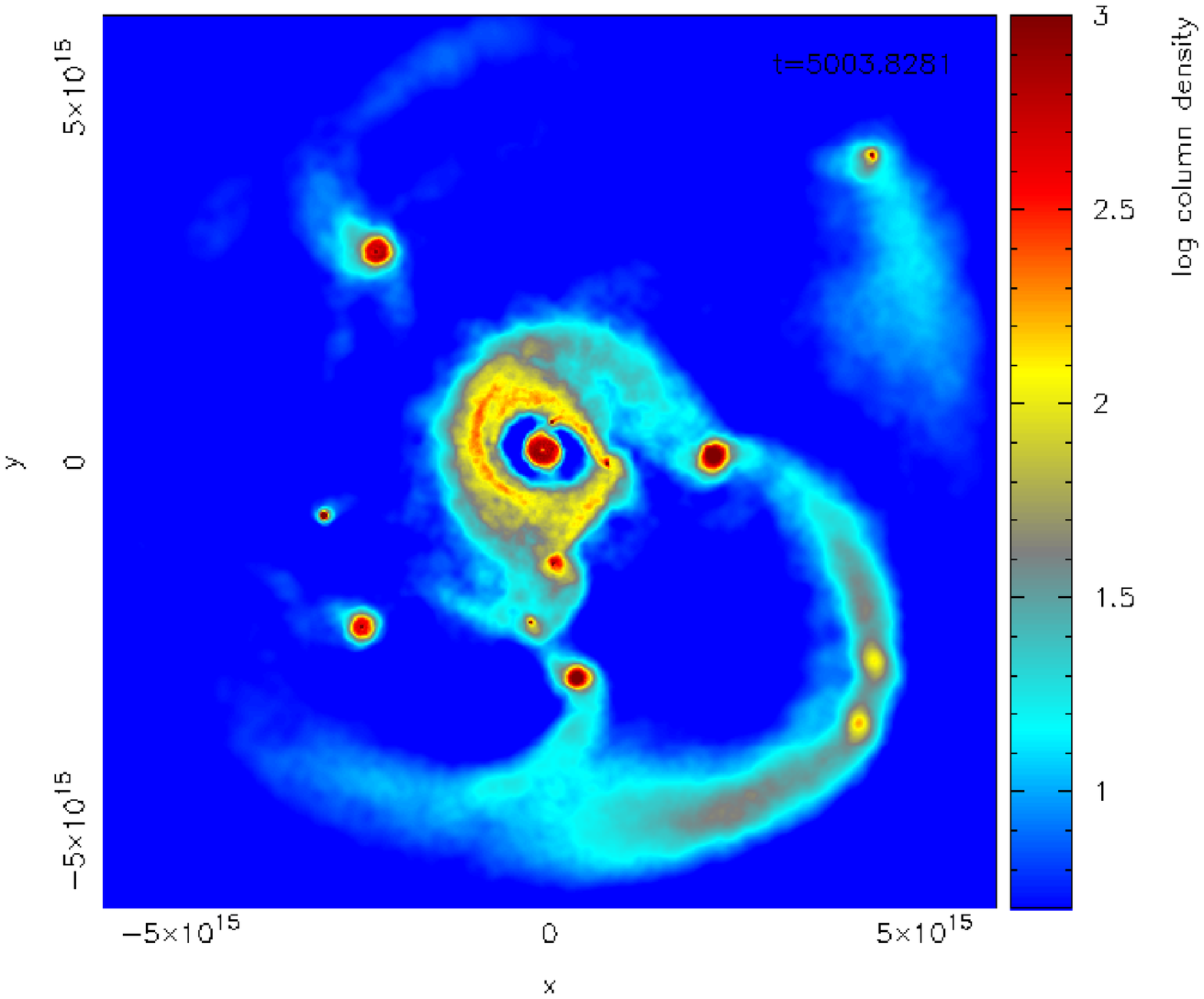}}
\caption{Radiative hydrodynamic simulation of the evolution
of a $0.5\,{\rm M}_\odot$ disc around a $0.7\,{\rm M}_\odot$ star.
Snapshots from 1000 to $5000\,{\rm yr}$ every $500\,{\rm yr}$ (as marked on each frame).
The disc is gravitationally unstable and can cool efficiently, hence
it quickly fragments to form 11 objects: low-mass H-burning stars, 
BDs, and planetary-mass objects. These objects form at radii from 
$\sim 100$ to $\sim 300\,{\rm AU}$ but due to mutual interactions a large 
fraction (9/11) escape.}
\label{fig:density}
\end{figure*}

\begin{figure}
\centerline{
\includegraphics[height=11cm]{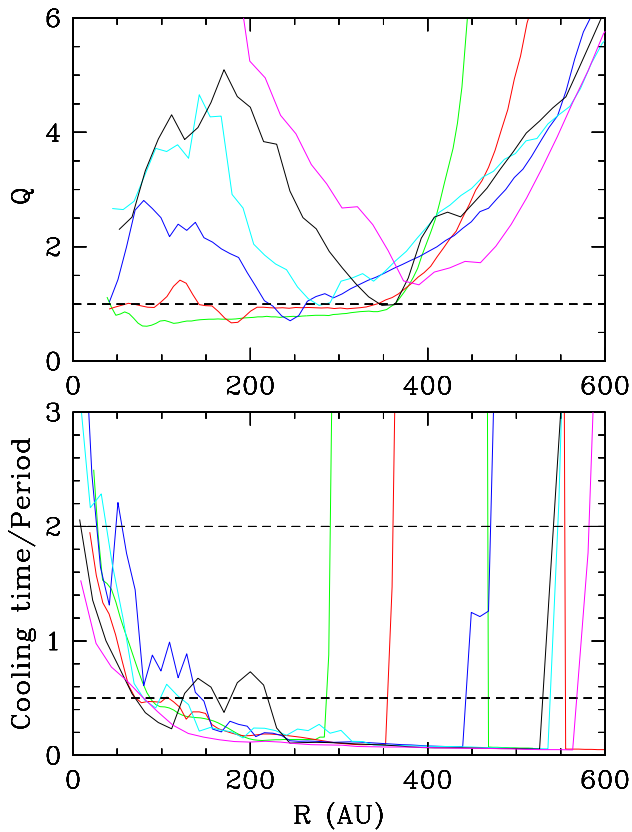}}
\caption{Toomre parameter Q (azimuthally averaged), and cooling time (in units of
the local orbital period), against radius at 6 different snapshots of the SPH simulation 
(from $1000\,{\rm yr}$ to $6000\,{\rm yr}$ every $1000\,{\rm yr}$; green, red, blue, cyan, black, magenta).
}
\label{fig:toomre}\label{fig:cool}
\end{figure}

\begin{table*}
\begin{minipage}{0.9\textwidth}
\caption{Properties of the objects produced by disc fragmentation
($t_i$: formation time; $R_i$ : radius at formation;
$m$: object mass at the end of the SPH simulation;
$R$, $v$: radius and velocity at the end of the SPH simulation;
$R_f$, $v_f$: radius and velocity at the end of the NBODY simulation).
}
\label{tab:fragments}
\centering
\renewcommand{\footnoterule}{}  
\begin{tabular}{@{}cccccccl} \hline
\noalign{\smallskip}
 $t_i$ (yr)
& $R_i$ (AU)
& $m$ (M$_{\sun}$) 
&  $R$ (AU) 
& $v$ (km s$^{-1}$) 
&  $R_f$ (AU) 
& $v_f$ (km s$^{-1}$)
& Comments \\
\noalign{\smallskip}
\hline
\noalign{\smallskip}
   0     &   0      &  0.712  &    0  & 0     &      0   &  0   & central star \\
   3337  &   110    &  0.058  &   16  & 5.9   &   9296   &  1.2 & BD, ejected in a  BD-BD pair  \\
   4052  &   118    &  0.079  &  451  & 1.8   &  43380   &  0.6 & BD, ejected in a  BD-BD pair\\
   4074  &    44    &  0.086  &   77  & 3.0   &      8   & 10.5 & low-mass star, bound  \\
   4132  &   327    &  0.036  &  464  & 0.7   &  43383   &  4.2 & BD, ejected in a BD-BD pair \\
   4242  &   187    &  0.025  & 1618  & 2.1   &  82214   &  1.7 & BD, ejected \\
   4346  &   286    &  0.035  &  191  & 2.7   &   9293   &  2.8 & BD, ejected in a BD-BD pair  \\
   4385  &   245    &  0.036  &  193  & 0.8   &    208   &  1.9 & BD, bound \\
   5495  &   159    &  0.027  & 2655  & 2.1   &  97448   &  2.1 & BD, ejected \\
   5540  &    94    &  0.015  &  135  & 1.2   & 204540   &  3.5 & BD, ejected \\
   5594  &   335    &  0.010  &  794  & 1.0   & 745941   & 12.7 & planetary-mass, ejected \\
10530    &   372   &   0.005  &  498  & 1.6   &  12064   &  1.1 & planetary-mass, ejected  \\
\noalign{\smallskip}
\hline
\end{tabular}
\end{minipage}
\end{table*}

\begin{figure}
\centerline{
\includegraphics[height=8.3cm,angle=-90]{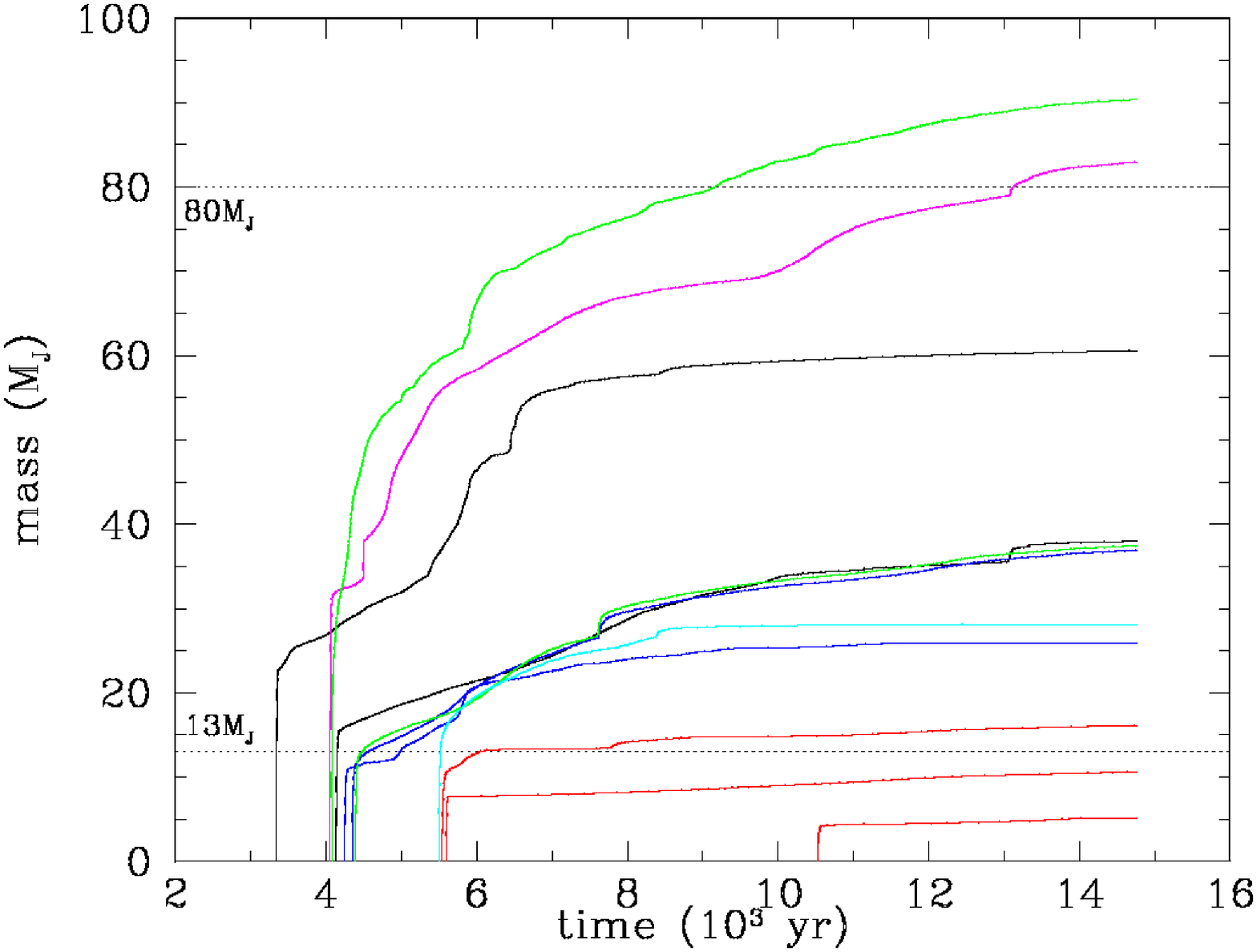}}
\caption{Fragment masses against time. The dotted lines correspond 
to the notional divisions between BDs and planets ($13\,{\rm M}_{_{\rm J}}$), and between BDs and stars ($80\,{\rm M}_{_{\rm J}}$).}
\label{fig:fmass}
\centerline{
\includegraphics[height=8.3cm,angle=-90]{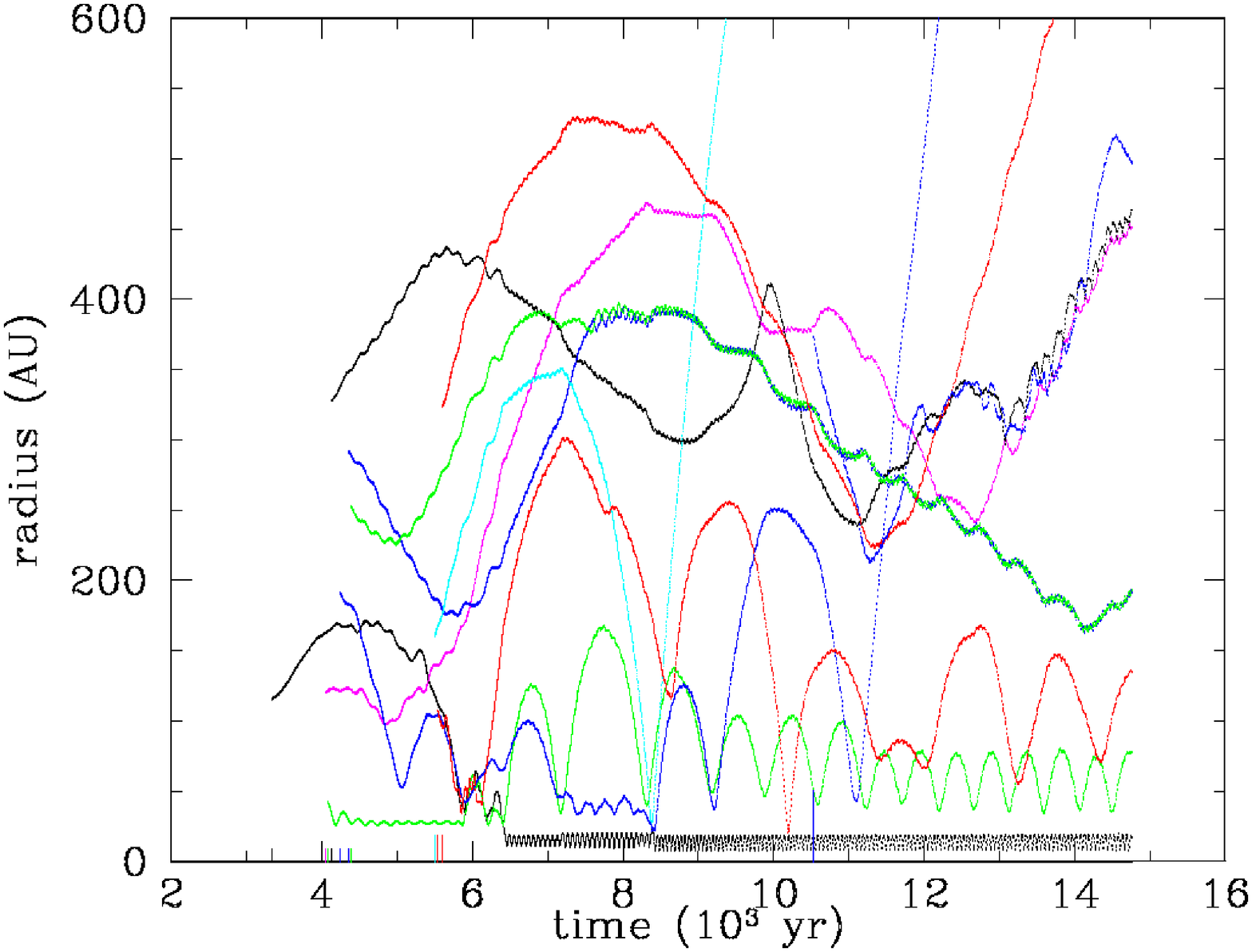}}
\caption{Orbital radius of the objects produced by disc
fragmentation against time. 10 out of 11 objects form at $>100\,{\rm AU}$, 
but they are then scattered by encounters with other objects in the disc.}
\label{fig:fradius}
\end{figure}

\begin{figure}
\centerline{
\includegraphics[height=8.3cm,angle=-90]{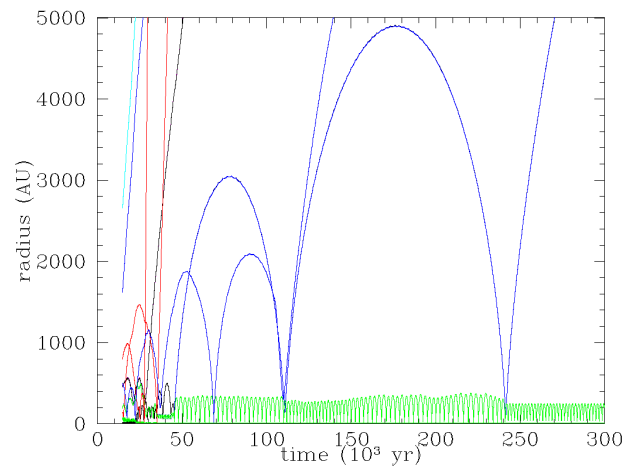}}
\caption{Long term evolution of the system using an $N$-body code. 
Orbital radius against time for all 11 objects. By $0.3\,{\rm Myr}$
only 2 objects remain bound to the central star.}
\label{fig:bdisc2mass}
\end{figure}

\section{Discussion}

The most widely discussed mechanisms for forming brown dwarfs (Whitworth et al. 2007) are `embryo ejection' from a forming small-$N$ proto-cluster (Reipurth \& Clarke 2001), and `turbulent fragmentation' (Padoan \& Nordlund 2004).

However, simulations of `embryo ejection' tend to produce BDs with unacceptably high velocity dispersions, insufficient discs to sustain accretion and outflows, and low binary fraction. It also relies on the notion that protostars grow from very low-mass seeds which, at their inception, are sufficiently compact to undergo essentially $N$-body dynamics.

Simulations of `turbulent fragmentation' have not yet addressed satisfactorily the clustering properties of young stars (e.g. the tendency for massive stars to form at the centres of dense star clusters); the binary statistics of young stars (e.g. the steady decrease in binary fraction and mean semi-major axis with decreasing primary mass); the systematic differences to be expected between the stellar IMF and the observed core mass function (e.g. when account is taken of failed prestellar cores and the variance in core lifetimes); and the failure of observers to detect the predicted velocity fields at the boundaries of prestellar cores.

Disc fragmentation appears to be a viable alternative scenario for forming BDs:

\begin{itemize} 

\item We have shown that massive, extended discs will sometimes form around Sun-like stars, and that these discs will then fragment producing multiple low-mass companions, principally BDs, but also planetary-mass objects and low-mass H-burning stars. This result is demonstrated here by means of numerical simulations, taking proper account of the energy equation and associated radiative transfer effects. It has also been derived analytically by Whitworth \& Stamatellos (2006).

\item BDs formed in this way are readily liberated into the field by interactions amongst themselves. We use the term `liberated' to emphasize that this is a more gentle process than `ejection'.

\item Close BD/BD binaries are quite a common outcome of disc fragmentation, and some survive liberation.

\item Liberated BDs frequently retain their own discs, with sufficient mass to sustain accretion and outflows.

\item The ratio of BDs to H-burning stars should be only weakly dependent on environmental factors, since it is largely controlled by the local physics of massive extended circumstellar discs.

\item The BDs formed here are not embryonic; they mop up most of the matter in the outer disc before being liberated.

\item Since one disc can spawn a large number of brown dwarfs (in this simulation 8), it may only be necessary for $\sim 10\%$ of Sun-like stars to undergo the process modelled here to supply all the observed brown dwarfs.

\item Planetary-mass objects are also formed by this mechanism (in this simulation 2) and are subsequently
liberated into the field to become free-floating objects (e.g. Lucas \& Roche 2000).

\item This mechanism also provides a way of forming solar-mass binary systems with low mass ratios, $q \la 0.2$ (cf. Delgado Donate \& Clarke 2005).

\item  We are not proposing a clear distinction between the way BDs form and the way H-burning stars form. Rather, as one progresses to lower-mass objects (from low-mass H-burning stars to BDs and then planetary-mass objects), the mix between those formed as primary protostars at the centres of collapsing prestellar cores, and those formed as secondary protostars by disc fragmentation, shifts in favour of the latter. This shift contributes to the downturn in the IMF at low masses, as the number of primaries decreases with decreasing mass.

\end{itemize}

The simulation presented here should be perceived purely as a thermodynamic theorem pertaining to an idealised situation, a `proof of concept'. The disc we invoke is so unstable that it would start to fragment whilst it was forming, and hence before it could relax to the circularly symmetric initial equilibrium we have used. Simulations with more dynamic initial conditions are needed to address this issue, and will be undertaken in future.

\section*{Acknowledgements}
   
The computations reported here were performed using the UK Astrophysical Fluids Facility (UKAFF). We also acknowledge support by PPARC grant PP/E000967/1. The final draft benefited significantly from the comments of the referee, Dr. Paolo Padoan.

\label{lastpage}

\end{document}